\documentclass[11pt]{article}
\usepackage{mymoriond,epsfig}

\def\Journal#1#2#3#4{{#1} {\bf #2}, #3 (#4)}

\def\NPB{{\em Nucl.\ Phys.\ }B}
\def\PLB{{\em Phys.\ Lett.\ } B}
\def\PRL{\em Phys.\ Rev.\ Lett.\ }
\def\PRD{{\em Phys.\ Rev.\ }D}

\def\IJMP{{\em Int.\ J.\ Mod.\ Phys.\ }A}

\def\be{\begin{equation}}
\def\ee{\end{equation}}
\def\bea{\begin{eqnarray}}
\def\eea{\end{eqnarray}}

\begin{document}
\vspace*{4cm}
\title{CURRENT TOPICS IN B PHYSICS$\:$\protect{\footnote{Plenary Talk
      given at IVth Rencontres de Vietnam, Hanoi, July 2000.}}}

\author{PATRICIA BALL}

\address{CERN/TH\\1211 Geneva 23, Switzerland}

\maketitle\abstracts{I discuss topics of particular interest in
  connection with the ongoing experiments at the B factories BaBar and
  Belle: B-mixing, QCD effects in nonleptonic decays and rare decays.}

\section{Introduction}

We currently witness the dawn of a very exciting era for B
physics: after years of planning and construction, the two B factories
BaBar and Belle have finally started operating in 1999 and first
thrilling results have been presented at this year's ICHEP 2000
conference \cite{BaBe}. The physics programme of the B factories
has two clear focal points: the detailed exploration of CP violation
in $B_d$ decays on the one hand and the precise measurement of
rare flavour-changing neutral current (FCNC)
processes on the other hand, with the aim to find or at least
constrain new physics. In the near and intermediate future, the
investigation of B decays will also form an essential part of the
physics programme of Tevatron Run II \cite{run2} and the LHC.\cite{LHC}

Why these enormous efforts? --- The answer to this question can maybe be
formulated as: ``Because B physics probes the scalar sector of the
Standard Model.'' Recall that the phenomena of 
both quark mixing, i.e.\ the CKM-matrix, and
CP-violation are inseparably connected to the fact that quarks have
mass; indeed, CP is a manifest and natural symmetry of massless gauge field
theories, both chiral and vector-like:$\:$\footnote{As is well known, also
  strong CP-violation -- the $\theta$ term in the Lagrangian -- is not
  observable if at least one quark is massless.} CP-violation is hence
intrinsically connected to the mechanism that gives mass to
particles, i.e.\ the Higgs-mechanism in the SM. 
The information on the scalar sector that can be obtained
from B decays thus has to be viewed as complementary to that
from direct Higgs-searches, and a complete picture of the mechanism of
SU$_{\rm L}$(2)$\times$U(1) symmetry-breaking and mass-generation
can only be
obtained by putting together the information from both direct searches and
B physics experiments. The reason why it is just the B system that is
so well suited for such studies
can be traced back to the fact that the top
is much heavier than the other quarks, which relaxes the 
GIM-suppression effective in box and
penguin-diagrams with near-degenerate quarks. The relaxation of
the GIM-mechanism also entails a
large $B_d$ mixing-phase, which gives rise to sizeable CP-violating
effects in B decays, and yields
typical branching ratios of ``rare'' FCNC $b\to s$ transitions of the
order 10$^{-6}$, which is by orders of magnitude larger than in rare
K and D decays.

Rare decays also offer the possibility to find or constrain
new-physics effects -- the experimental results for the
penguin-mediated process $b\to s\gamma$, for instance, serve as severe
constraints of sfermion masses and mixing in SUSY-models. Also here
one observes a certain complementarity between direct searches and B
physics: once new physics, e.g.\ SUSY, is found at the Tevatron or the
LHC, it will entail simultaneous production of a plethora of new
particles, making it hard to impossible to disentangle their
decay chains and determine particle parameters in a model-independent way.
Also here
B physics will help in putting constraints from the observed indirect
effects of such new particles and thus fill the gap between new-physics
discovery at hadron machines and the
takeoff of a linear collider at which the properties of new particles
can be studied in detail.

The above considerations let it appear as approriate to center this 
overview around theoretical challenges in CP-violating processes and 
rare decays; for the dicussion of other, also
very interesting topics in B physics, like e.g.\ semileptonic $b$
decays, I refer to 
  Bigi's talk at ICHEP 2000.\cite{bigi}

\section{B-Mixing}

\begin{figure}
\centerline{\epsfxsize=0.8\textwidth\epsffile{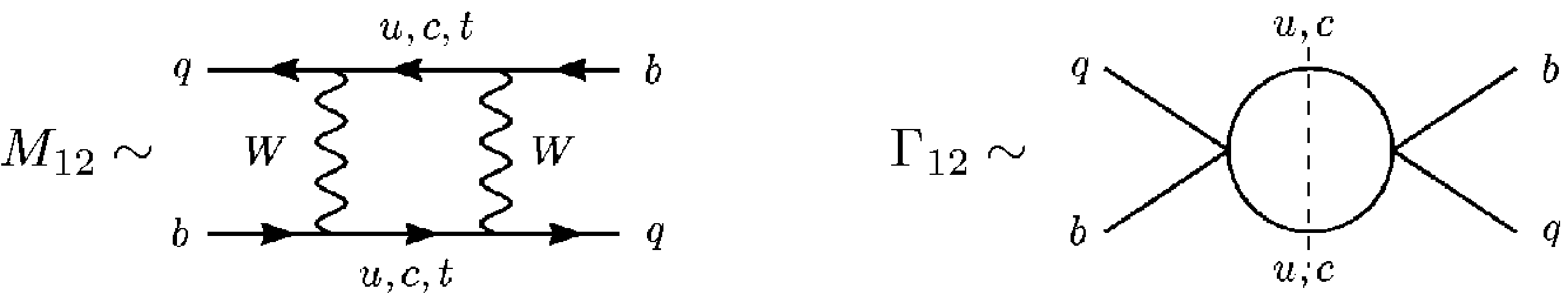}}
\caption[]{Typical diagrams contributing to B-mixing. The dashed line
  in the diagram to the right denotes that the imaginary part is to be
  taken. Figure taken from Ref.\protect{\cite{LHC}}.}\label{fig:1}
\end{figure}

Why start this review with a section on B-mixing? --- Because it is
highly relevant both for understanding CP-violation and for the
measurement of the CKM matrix elements $|V_{tq}|$ with
$q=d,s$. Let me
shortly review the essentials: As is
well known, the flavour-eigenstates of neutral B mesons, $B^0_q =
(q\bar b)$ and $\bar B^0_q = (b\bar q)$, mix on account of weak
interactions. The mixing can be described, in the framework of 
quantum mechanics and in the basis of flavour-eigenstates, by the Hamiltonian
$$
{\bf H} = {\bf M} - \frac{i}{2}\,{\bf \Gamma} =
\left( \begin{array}{cc} M & M_{12}\\ M_{12}^* & M \end{array} \right)
- \frac{i}{2} \left( \begin{array}{cc} \Gamma & \Gamma_{12}\\ 
\Gamma_{12}^* & \Gamma \end{array} \right),
$$
where ${\bf M}$ and ${\bf \Gamma}$ are Hermitian and their
respective diagonal entries are equal by virtue of CPT
invariance. Typical diagrams contributing to the off-diagonal elements
are shown in Fig.~\ref{fig:1}.
The induced mass and width differences of
the mass eigenstates, i.e.\  the observed B mesons, are given by
$$
\Delta M_q = 2 \left| M_{12}^{(q)} \right|, \qquad \Delta \Gamma_q =
2\, \frac{{\rm Re}\, (M_{12}^{(q)*}
  \Gamma_{12}^{(q)})}{|M_{12}^{(q)}|}\,.
$$
A quantity especially relevant for CP measurements is the mixing-phase
$\phi_q = \arg\, M_{12}^{(q)}$. In experiment, $\Delta M_d$ has been
measured with small error, but for $\Delta M_s$
there exists only a lower bound to date; in the SM, the expected value is $\sim
(15-25)\,$ps$^{-1}$, which induces $B_s$ oscillations too fast to be 
resolved at present machines, but is well within the reach of Tevatron
Run II and the LHC. The width-difference in the $B_d$
system is expected to be too small to be measurable, but could be
non-negligible for $B_s$; present experimental results for the latter
are too crude to be conclusive.

As the theoretical expression for $\Delta M_d$ is directly
proportional to $|V_{td}|^2$, the sum over 
box-diagrams being dominated by the $t$ quark contribution, one would
think that a precise measurement of $|V_{td}|$ be
possible. The achievable precision does, however, crucially depend on
the accuracy to which the relevant hadronic matrix element, $\langle B^0_d
\mid (\bar d b)_{V-A} (\bar d b)_{V-A} \mid \bar B^0_d \rangle\sim
f_B^2 \hat B_{B_d}$, is
known. Best available numbers come from lattice simulations; for a
comprehensive review of the
state of the art of lattice simulations of B meson matrix elements I
refer to \cite{wittig}; here, let it suffice to say that simulating the
$b$ quark with its mass $\sim 5\,$GeV, i.e.\ a Compton-wavelength not
large compared to typical lattice spacings, $a\sim (2-4)\,$GeV$^{-1}$,
poses a severe challenge, which, together with the fact that most
simulations are done in the quenched approximation, neglecting the
feedback of quarks on the gauge-fields, entails a large
quoted error of $\sim\,$30\% on $f_B^2 \hat B_{B_d}$. Preliminary
results from unquenched
simulations have been presented by the CP--PACS collaboration at ICHEP
2000$\:$\cite{CPPACS}; they indicate an increase of $f_B$ by ca.\ 10\%
with respect to the quenched results.

A quantity in which much of lattice systematics and in particular
quenching effets are expected to cancel, is
the ratio of $B_s$ and $B_d$ matrix elements, $(f_{B_s} \sqrt{\hat
B_{B_s}})/(f_{B_d} \sqrt{\hat B_{B_d}}) \equiv \xi$, quoted as
$1.16\pm 0.07$ \cite{kenway}, which allows one to relate $\Delta
M_s/\Delta M_d$ to the ratio of CKM matrix elements $|V_{ts}/V_{td}|$ as 
$$
\frac{\Delta M_s}{\Delta M_d} = \left|\frac{V_{ts}}{V_{td}}\right|^2
\frac{M_{B_s}}{M_{B_d}} \, \xi^2\,.
$$
Thus, a determination of $|V_{ts}/V_{td}|$ will be possible with small
theoretical error once $\Delta M_s$ has been measured. Note that the
above formula is valid within in the SM only and can be upset by 
new-physics contributions to B-mixing; 
in this case, one expects this determination of
$|V_{ts}/V_{td}|$ to be at variance both with the one from
FCNC $b\to s$ and $b\to d$ transitions and with
constraints from the unitarity of the CKM matrix.

\section{Challenges I: Nonleptonic Decays \& CP-Violation}

Measuring CP-violating time-dependent asymmetries is at the heart of
BaBar's and Belle's physics programme. CP-asymmetries are
in general non-vanishing only because of the presence of strong phases ---
and it is just the same strong phases that often complicate the aim of such 
measurements, i.e.\ the
extraction of CP-violating weak phases.
To illustrate the issue, let us consider the
decay of $B_d$ into a final state $F$ which is an eigenstate under
CP, CP$\mid F\rangle = \pm \mid F\rangle$, such as $J/\psi K_S$ or
$\pi\pi$. In this case, the time-dependent CP-asymmetry can be written
as
\begin{eqnarray}
a_{\rm CP}(t) & = & \frac{\Gamma(B^0_d(t)\to F)-\Gamma(\bar
  B^0_q(t)\to F)}{\Gamma(B^0_d(t)\to F)+\Gamma(\bar
  B^0_q(t)\to F)}\nonumber\\
& = & {\cal A}_{\rm CP}^{\rm dir}(B_d\to F) \cos (\Delta M_d t ) + 
{\cal A}_{\rm CP}^{\rm mix}(B_d\to F) \sin (\Delta M_d t
  )\,,\label{eq:Acp}
\end{eqnarray}
where the observables ${\cal A}_{\rm CP}^{\rm dir,mix}$ can be
expressed in terms of the hadronic quantity
\begin{equation}
\xi_F = \mp e^{-i\phi_d} \,\frac{\langle F \mid H^{\rm eff}_{\rm weak}
  \mid \bar  B^0_d\rangle}{\langle F \mid H^{\rm eff}_{\rm weak} \mid 
B^0_d\rangle}
\end{equation}
as
\begin{equation}
{\cal A}_{\rm CP}^{\rm dir} = \frac{1-\left|\xi_F\right|^2}{1+
\left|\xi_F\right|^2}, \qquad {\cal A}_{\rm CP}^{\rm mix} = \frac{2
{\rm Im}\,\xi_F}{1+\left|\xi_F\right|^2}\,;
\end{equation}
$\phi_d = \arg M_{12}$ is the $B_d$ mixing-phase introduced in the
previous section and 
$$
H^{\rm eff}_{\rm weak}=\sum_{j=u,c,t; r=s,d} V^*_{jr}
V^{\vphantom{*}}_{jb} Q^{jr}+ \mbox{c.c.}
$$
the effective weak Hamiltonian, written as sum over weak amplitudes. 
As $\xi_F$ in general depends on hadronic matrix
elements, its exact calculation is possible only for the
special case of one single weak amplitude dominating, such that the
hadronic matrix elements in $\xi_F$ cancel; these are the so-called
``gold-plated'' decays \cite{BS}, e.g.\ $B_d\to J/\psi K_S$, where to
good approximation $\xi_F$ is a pure phase. $B_d\to J/\psi K_S$ is
also to date the only B decay channel where CP-violation has been
observed. The results of BaBar and Belle may give first hints at a
non-standard mechanism of CP-violation, with --- amusingly ---
numerical results not too
far away from predictions in a model where CP-violation is understood
as spontaneous breakdown of a manifest CP-symmetry of the underlying 
Lagrangian.\cite{SBLR} However, large experimental
uncertainties do not yet allow a definite conclusion. 

The majority of data,
however, will be collected in ``brazen'' channels, where the
extraction of weak phases requires precise knowledge of hadronic
matrix elements, i.e.\ nonperturbative QCD. As is well known, in the
SM, CP-violation is due to one single complex phase in the CKM-matrix;
the discussion in the literature, however, often refers to three 
``weak phases'', labelled $\alpha$, $\beta$, $\gamma$, which are
the three angles of the ``unitarity triangle'', i.e.\ the graphic
representation of the unitarity relation $\sum_{q=u,c,t} V_{qb}
V_{qd}^* = 0$ in the complex plane; with the usual phase-conventions
of the CKM-matrix one has e.g.\ for the B-mixing phases 
$\phi_d=2\beta$ and $\phi_s\approx 0$.
The aim of CP-measurements is then
to overconstrain the unitarity triangle by measuring its sides and
angles in as many ways as possible in order to verify or falsify the
CKM-picture of CP-violation.

I would like to stress here that the problem of how to calculate
nonleptonic B hadronic matrix elements with phenomenologically
acceptable precision is indeed very important and that despite recent
progress to be reported below it is premature to consider the problem as
sufficiently well understood. Much work is still needed before we can
extract weak phases from ``brazen'' channels with a similar accuracy
as from the ``gold-plated'' ones.\footnote{Note that in contrast to 
B-mixing, lattice calculations are not likely to provide substantial help
   in the foreseeable future, as the simulation of a
  two-particle final state with sharp momentum requires a lattice with
  spatial extension by orders of magnitude larger than what is within
  present reach.} 

\subsection{``Diagrammatics'' and U-Spin Flavour-Symmetry}

One rather pragmatic possibility to treat the unknown 
QCD matrix elements is to exploit dynamical symmetries of QCD in order
to reduce the number of independent matrix elements and actually {\em
  measure} them in experiment rather than calculate them from first
principles. A representative example for this approach, discussed in
Ref.~\cite{fleischer}, is provided by
the pair of decay channels $B_d\to\pi^+\pi^-$ and $B_s\to K^+ K^-$
which are related by U-spin symmetry, i.e.\ the exchange of $d$ and
$s$ quarks, under which QCD is assumed to be invariant. 
The decay amplitudes can be written as
\begin{eqnarray}
A(B_d\to\pi^+\pi^-) & \propto & e^{i\gamma} \left[ 1 - d e^{i\theta}
  e^{-i\gamma} \right],\nonumber\\
A(B_s\to K^+K^-) &\propto& e^{i\gamma} \left[ 1 + \mbox{(CKM
  factor)}\,\times\, d' 
e^{i\theta'} e^{-i\gamma} \right],\label{eq:ampl}
\end{eqnarray}
where the terms in $d e^{i\theta}$ and $d' e^{i\theta'}$ denote the
disturbing ``penguin-pollution'' contributions. In the limit of
U-spin symmetry be
realized exactly, one has $d=d'$ and $\theta=\theta'$, and the four
CP-observables accessible in the two channels can be expressed in terms of
only three unknowns, the weak phase $\gamma$ and the two
penguin-parameters $d$ and $\theta$, provided the mixing-phases
$\phi_d$ and $\phi_s$ have been measured in gold-plated channels
before. The validity of this approach is, however, limited by U-spin
breaking effects, which need not necessarily be small. We have already
encountered an example in the previous section:
$$
\frac{\langle B_s^0 \mid (\bar s b)_{V-A} (\bar s b)_{V-A} \mid \bar
  B_s^0\rangle}{\langle B_d^0 \mid (\bar d b)_{V-A} (\bar d b)_{V-A} \mid \bar
  B_d^0\rangle} = 1.37 \pm 0.16 \neq 1\,.
$$
Also the ratio of the two amplitudes given in (\ref{eq:ampl})
does, in factorization approximation, and setting all CKM factors
equal, deviate from 1:$\:$\cite{FF}
$$
\frac{A(B_s\to K^+K^-)}{A(B_d\to\pi^+\pi^-)} = \frac{f_K}{f_\pi}\,
\frac{F_0^{B_s\to K}(M_K^2)}{F_0^{B_d\to\pi}(M_\pi^2)} \left(
  \frac{M_{B_s}^2 - M_K^2}{M_{B_d}^2 - M_\pi^2} \right) \approx 1.35.
$$
Once experimental data are available, the accuracy to which U-spin symmetry
is realized in (\ref{eq:ampl}) can partially be tested by
relaxing one of the two conditions $d=d'$ and $\theta=\theta'$, so
that one extracts four unknowns from four
observables. Further theoretical studies of 
U-spin breaking effects are, however,
definitely needed before one can use U-spin relations with confidence as a
precision tool to extract weak phases.

\subsection{Hard Perturbative QCD and the Heavy Quark Limit}

In the limit $m_b\to\infty$, nonleptonic B decays share some crucial
features with hard exclusive QCD reactions at large momentum transfer,
such as e.g.\ hadron electromagnetic form factors. The theoretical
description of such processes was pioneered by Brodsky and Lepage (see
e.g.~\cite{BLreport}) who showed that, to leading order in a
light-cone expansion in terms of contributions of increasing twist,
i.e.\ in terms of contributions of increasing inverse power of
momentum-transfer, factorization is possible and the amplitude can be
written as convolution of a
process-dependent, perturbative hard-scattering kernel with
 universal, nonperturbative functions describing the
parton momentum-distribution inside the hadron; the
factorization is such that only parton-momenta in direction of the
particle momentum do contribute and transverse degrees of freedom are
suppressed -- hence the
term ``collinear factorization''. Recently, Beneke et al.\ have shown 
that a conceptually similar approach can also be applied to
nonleptonic B meson decays, where, loosely speaking, the $b$ quark
mass plays the r\^{o}le of the large parameter and serves to suppress
higher-twist contributions. To O$(\alpha_s)$, the amplitude of the
process $B\to\pi\pi$ can be written as~\cite{BBNS}
\begin{eqnarray}
\langle \pi\pi \mid H_{\rm weak}^{\rm eff} \mid B\rangle & = &
f^{B\to\pi}_+(0) \int_0^1 dx\, T^I(x) \,\phi_\pi(x)\nonumber\\
&&{} + \int_0^1
d\xi\,dx\,dy\, T^{II}(\xi,x,y)\, \phi_B(\xi) \phi_\pi(x)\phi_\pi(y) +
O(1/m_b).\label{eq:4}
\end{eqnarray}
\begin{figure}
\centerline{\epsfxsize=0.6\textwidth\epsffile{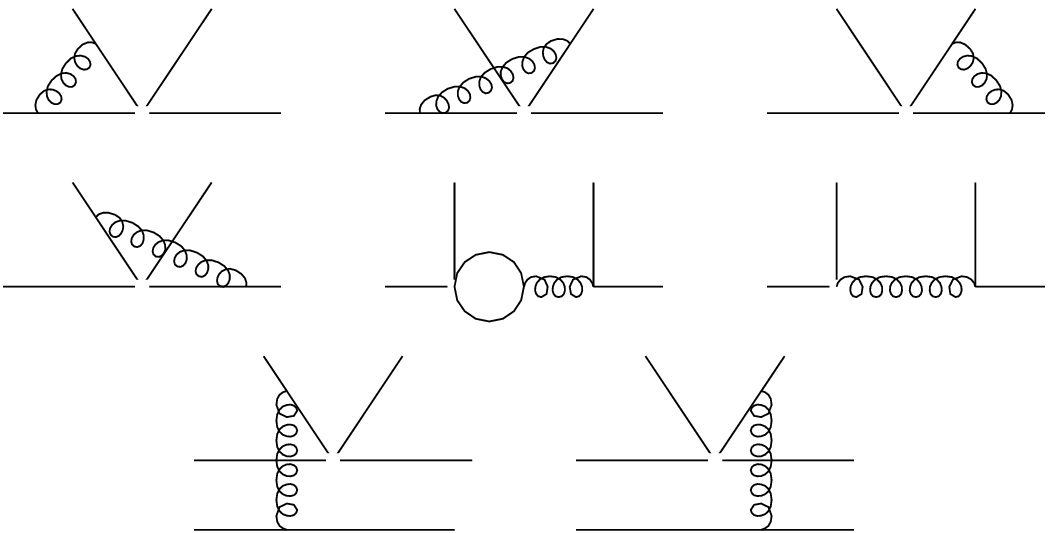}}
\caption[]{O($\alpha_s$) corrections to the hard-scattering kernels
  $T^I$ (first two rows) and $T^{II}$ (last row). The two lines
  directed upwards represent the two quarks forming the emitted
  pion. Figure taken from Ref.\cite{BBNS}.}\label{fig:2}
\end{figure}
The diagrams contributing to the hard-scattering kernels are shown in
Fig.~\ref{fig:2}. $\phi_\pi$, technically speaking the twist-2
distribution amplitude of the $\pi$, is a rather well-studied object
whose functional dependence on $x$ can be understood, exploiting
conformal symmetry of massless QCD (symmetry-group SL(2,R)), in terms of a
partial wave expansion in terms of contributions of increasing
conformal spin.\cite{BLreport,BBKT} Much less, however, is known about
the B meson's distribution amplitude $\phi_B$, for whose
parametrization one has to rely on models.
The numerical analysis of (\ref{eq:4}) reveals that the
``penguin-pollution'' term
$d\exp(i\theta)$ of the last section is small and that the branching
ratio reads
\begin{equation}
B(\bar B_d \to \pi^+\pi^-) = 6.5\cdot 10^{-6}\times \left| e^{-i\gamma} +
  0.09\, e^{i\cdot 12.7^\circ} \right|^2,\label{eq:5}
\end{equation}
which indicates that corrections to the time-honoured 
``vacuum-saturation'' picture are small.

The crucial question is now about the size of power-suppressed $1/m_b$
corrections to (\ref{eq:4}).$\,$\footnote{Note also that even in the
  strict limit $m_b\to\infty$, collinear factorization may fail in the
  second term on the 
right-hand side of (\ref{eq:4}) at O$(\alpha_s^2)$.$\,$\cite{defeat}} 
As discussed in \cite{defeat}, there are
two problems associated with them: first, there are formally
power-suppressed, but chirally enhanced
twist-3 terms in $2m_\pi^2/(m_u+m_d)/m_b\approx 0.7$, which is not a small
parameter. Second, at O$(\alpha_s)$ these terms induce logarithmic
divergences which violate the QCD factorization formula
(\ref{eq:4}). On a more technical level, these logarithmic
  divergences are due to soft contributions similar to those that, 
  in principle, also appear at leading order in
  $1/m_b$, but in (\ref{eq:4}) are subsumed in the form factor 
  $f_+^{B\to\pi}$. Soft terms actually spoil the calculation of B
  decay form factors by conventional hard perturbative QCD
  methods,$\:$\footnote{They are, however, included in a systematic
    way in a conceptionally similar, alternative approach for
    calculating heavy-to-light form factors: QCD sum
    rules on the light-cone.$\,$\cite{LC}} and it is probably the main
  achievement of~\cite{BBNS} as compared to
  competing approaches~\cite{li}, also based on hard
  perturbative QCD, to show that, for nonleptonic B decays, at leading
  order in $1/m_b$, all soft terms can be included into the
  experimental observable $f_+^{B\to\pi}(0)$. Nevertheless it is
  somewhat disturbing that soft effects show up again at order
  $1/m_b$ and are chirally enhanced.

The emerging picture thus seems to be that the formal
limit $m_b\to\infty$ can be treated on theoretically safe grounds, but
that power-suppressed terms in $1/m_b$, whose treatment in a systemtic
way is not yet understood, are phenomenologically relevant. I
note in passing that this appears to be 
a general feature of calculations in the
heavy quark limit: unless terms linear in $1/m_b$ are absent due to a
protecting symmetry, they yield sizeable contributions.\footnote{In
  this connection, I recall the
  long-standing 
  discussion of the (comparatively) simple case of $1/m_b$ corrections to
  the leptonic decay constant of the B meson, $f_B$; cf.\
  Ref.~\cite{fB} and references therein.}
It thus appears presently very difficult to attach any meaningful
theoretical uncertainty to (\ref{eq:4}) and (\ref{eq:5}).

\section{Challenges II: Rare Decays}

Flavour-changing neutral current decays involving $b\to s$ or $b\to d$
transitions occur only at loop-level in the SM, come with small
exclusive branching ratios $\sim $O$(10^{-6})$ or smaller and thus
provide an excellent probe of indirect effects of new physics and
information on the masses and couplings of the virtual SM or
beyond-the-SM particles participating. Within the SM, these decays are
sensitive to the CKM matrix elements $|V_{ts}|$ and $|V_{td}|$,
respectively; a measurement of these parameters or their ratio would
be complementary to their determination from B-mixing.

The effective field theory
for $b\to s(d)$ transitions is universal for all channels;
due to space-restrictions, I cannot review all
important features of that effective theory; for a quick overview I
refer to Chapter~9 of the BaBar Physics Book \cite{BaBar}, where also
references to more detailed reviews can be found. Here, let it suffice
to say
that the effective Hamiltonian governing rare decays can be obtained
from the SM Hamiltonian by performing an operator product expansion
yielding
\begin{equation}\label{eq:hammel}
{\cal H}^q_{\mbox{\scriptsize eff}} = -\frac{4 G_F}{\sqrt{2}}\,
V_{tb}^{\vphantom{*}}
V_{tq}^* \,\sum\limits_{i=1}^{11} C_i(\mu) {\cal O}^q_i(\mu),
\end{equation}
where the ${\cal O}^q_i$ are local renormalized operators.
The Wilson-coefficients
$C_i$ can be calculated in perturbation theory and encode the
relevant short-distance physics, in particular any potential
new-physics effects; they are known to NLO in QCD.$\,$\cite{NLO}
The renormalization-scale $\mu$ can be viewed as
separating the
long- and short-distance regimes. For calculating decay rates with
the help of (\ref{eq:hammel}), the value of $\mu$ has to be chosen as
$\mu\sim m_b$ in a truncated perturbative expansion.
The Hamiltonian (\ref{eq:hammel}) is suitable to describe physics in the
SM as well as in a number of its extensions, for instance the minimal
supersymmetric model. The operator basis in (\ref{eq:hammel})
is, however, not always complete, and in some models, for instance
those exhibiting left-right symmetry, new physics also shows up in the
form of new operators. This proviso should be kept in mind when
analysing rare B decays for new-physics effects by measuring
Wilson-coefficients.

The challenge in rare decays is to correctly assess the size of
long-distance QCD contributions. Contrary to naive expectations, such
contributions do not only affect exclusive, but also inclusive
decays; for $b\to s$ transitions, such long-distance effects come from
short-distance $b\to c\bar c s$ transitions, where the $c\bar c$ pair
(plus soft and/or hard gluons) forms an intermediate state that at
large distances couples to a photon or a lepton-pair.

Let me discuss the issue in more detail for the simplest case of the
exclusive decay $B\to K^*\gamma$, concentrating on
non-perturbative QCD effects.  For the treatment
of perturbative issues, in particular the reduction of
renormalization-scale dependence and remaining uncertainties, I refer
to \cite{misiak}.

The theoretical description of the $B\to K^*\gamma$ decay is quite
involved with regard to both long- and short-distance contributions.
In terms of the effective Hamiltonian (\ref{eq:hammel}), the decay
amplitude can be written as
\begin{equation}\label{eq:bsg}
{\cal A}(\bar{B}\to \bar{K}^*\gamma) = -\frac{4 G_F}{\sqrt{2}}\,
V_{tb}^{\vphantom{*}} V^*_{ts}
\,\langle \bar{K}^* \gamma | C_7 O_7 + i\epsilon^\mu \sum_{i\neq 7}
C_i \int d^4x
e^{iqx} T \{ j_\mu^{em}(x)O_i(0)\} | \bar{B}\rangle\,,
\end{equation}
where $j_\mu^{em}$ is the electromagnetic current and $\epsilon_\mu$ the
polarization vector of the photon. $O_7$ is the only operator
containing the photon field at tree-level:
\begin{equation}
O_7 = \frac{e}{16\pi^2}\, m_b \bar s \sigma_{\mu\nu} R b F^{\mu\nu}
\end{equation}
with $R=(1+\gamma_5)/2$. Other operators, the
second term in (\ref{eq:bsg}), contribute mainly closed fermion loops.  The
first complication is now that the first term in (\ref{eq:bsg})
depends on the regularization- and renormalization-scheme. For this
reason, one usually introduces a scheme-independent linear combination
of coefficients, called ``effective coefficient'' (see \cite{misiak}
and references therein): $$C_7^{\mbox{\scriptsize eff}}(\mu) = 
C_7(\mu) + \sum_{i=3}^6 y_iC_i(\mu),$$
where the numerical coefficients $y_i$ are given in
\cite{misiak}.

The current-current operators $
O_1 = (\bar s\gamma^\mu L b)(\bar c\gamma_\mu L c) $ and
$O_2 = (\bar s\gamma_\mu L c)(\bar c\gamma^\mu L b)$
give vanishing contribution to the perturbative $b \to s \gamma$
amplitude at one loop.  Thus, to leading logarithmic accuracy (LLA) in
QCD and neglecting long-distance contributions from $O_{1,2}$ to the
$b\bar{s}\gamma X$ Green's functions, the $\bar{B}\to \bar{K}^*\gamma$
amplitude is given by
\begin{equation}
{\cal A}_{O_7}^{\rm LLA}(\bar{B}\to \bar{K}^*\gamma) =
-\,\frac{4G_F}{\sqrt{2}}\, V_{tb}^{\vphantom{*}}
 V_{ts}^* C_7^{(0){\mbox{\scriptsize eff}}}
\langle \bar{K}^*\gamma | O_7 | \bar{B}\rangle.
\end{equation}
Here, $C_7^{(0){\mbox{\scriptsize eff}}}$ denotes the leading logarithmic
approximation to $C_7^{\mbox{\scriptsize eff}}$. The above expression
is, however,
not the end of the story, as the second term in (\ref{eq:bsg}) also
contains long-distance contributions. Some of them can be viewed as
the effect of virtual intermediate resonances $\bar{B}\to \bar{K}^*
V^*\to \bar{K}^*\gamma$.  The main effect comes from $c\bar c$
resonances and is contributed by the operators $O_1$ and $O_2$ in
(\ref{eq:bsg}). It is governed by the virtuality of $V^*$, which, for
a real photon, is just $-1/m_{V^*}^2\sim -1/4 m_c^2$. Such
power-suppressed long-distance effects $\sim 1/m_c^2$ are also present
in inclusive decays, which is actually the process for which 
they were discussed first.$\:$\cite{voloshin}
The first, and to date only, study for exclusive decays was done in
\cite{stoll}.  Technically, one performs an operator product expansion
of the correlation function in (\ref{eq:bsg}), with a soft
non-perturbative gluon being attached to the charm loop, resulting in
terms being parametrically suppressed by inverse powers of the charm
quark mass.  As pointed out in \cite{misiak}, although the power increases
for additional soft gluons, it is possible that contributions of
additional external hard gluons could remove the power-suppression.
This question is also relevant for inclusive decays and deserves
further study.

After inclusion of the power-suppressed terms $\sim 1/m_c^2$, the
$\bar{B}\to \bar{K}^*\gamma$ amplitude reads
\begin{equation} \label{eq:with.OF}
{\cal A}^{\rm LLA}(B\to \bar{K}^*\gamma) = -\,\frac{4G_F}{\sqrt{2}}\,
V_{tb} V_{ts}^*
\langle \bar{K}^*\gamma \mid C_7^{(0){\mbox{\scriptsize eff}}} O_7 +
\frac{1}{4m_c^2}\, C_2^{(0)} O_F \mid \bar{B}\rangle.
\end{equation}
Here, $O_F$ is the effective quark-quark-gluon operator obtained in
\cite{stoll}, which describes the leading non-perturbative
corrections. The two hadronic matrix elements can be described in
terms of three form factors, $T_1$, $L$ and $\tilde{L}$:
\begin{eqnarray}
\langle \bar{K}^*(p)\gamma|\bar s \sigma_{\mu\nu} q^\nu b
|\bar{B}(p_B)\rangle & = & i\epsilon_{\mu\nu\rho\sigma}\epsilon^{*\mu}_\gamma
\epsilon^{*\nu}_{K^*} p_B^\rho p^\sigma 2 T_1(0),\nonumber\\
\langle \bar{K}^*(p)\gamma|O_F|\bar{B}(p_B)\rangle & = &
\frac{e}{36\pi^2} \left[
  L(0) \epsilon_{\mu\nu\rho\sigma} \epsilon^{*\mu}_\gamma \epsilon^{*\nu}_{K^*}
  p_B^\rho p^\sigma\right.\nonumber\\
& & \left.{} + i \tilde{L}(0) \left\{ (\epsilon^*_{K^*} p_B)
(\epsilon^*_{\gamma} p_B) - \frac{1}{2}\, (\epsilon^*_{K^*}
\epsilon^*_{\gamma}) (m_B^2 -
m_{K^*}^2) \right\} \right].
\end{eqnarray}
The calculation of the above form factors requires genuinely
non-perturbative input. Available methods include, but do not exhaust,
lattice calculations and QCD sum rules. Again, a discussion of the
respective strengths and weaknesses of these approaches is beyond the
scope of this talk. Let it suffice to say that --- at least at
present --- lattice cannot reach the point $(p_B-p)^2=0$ relevant for $B\to
K^*\gamma$, and that QCD sum rules predict the relevant form factors
with an estimated 20\% uncertainty.$\:$\cite{stoll,BB}
Numerically, these corrections increase the decay rate by about 5 to 10\%.
After their inclusion, one obtains
$$
B(B\to K^*\gamma) 
 =  4.4\times 10^{-5}\,\times\, (1+8\%)
$$
for the central values of the QCD sum rule results, where the second term
in brackets denotes the shift of the result induced by $1/m_c^2$ terms.

Let me also spend a few words on the decay
$B\to\rho\gamma$. Although at first glance it might seem that its
structure is the same as for $B\to K^*\gamma$, this is actually
not the case: there are additional long-distance
contributions to $B\to\rho\gamma$,
which are CKM-suppressed for $B\to K^*\gamma$ and have been neglected
in the previous discussion; these contributions comprise
\begin{itemize}
\item weak annihilation mediated by $O^{u}_{1,2}$ with non-perturbative
  photon-emission from light quarks; these contributions are
  discussed in \cite{Brhog} and found to be of order 10\% at the
  amplitude level;
\item effects of virtual $u\bar u$ resonances ($\rho$,
  $\omega$,\dots); they are often said to be small, but actually have
  not been studied yet in a genuinely
  non-perturbative framework.
\end{itemize}
{}From the above open questions it is evident that further theoretical
work is needed before one can aim at an accurate experimental
determination of $|V_{ts}/V_{td}|$ from
$B(B\to\rho\gamma)$ and $B(B\to K^*\gamma)$. 

I would like to conclude the discussion of rare decays with $B\to
K^*\ell^+ \ell^-$. The
motivation for studying this decay is either, assuming the SM to be
correct, the measurement of $|V_{ts}|$, or the
search for manifestations of new physics in non-standard values of the
Wilson-coefficients. A very suitable observable for the latter purpose is the
forward-backward asymmetry which is independent of CKM matrix elements,
but, due to extremely small event numbers, only accessible at the
LHC. $B\to
K^*\mu^+\mu^-$ has the 
potential of high impact both
on SM physics and beyond (see e.g.\ Ref.~\cite{ali} for a recent discussion
of potentially large SUSY-effects).

In $b\to s\ell^+\ell^-$ transitions, 
the issue of intermediate $c\bar c$ resonances is even more relevant
than it is for $b\to s\gamma$, as they show up as observable peaks in
the dilepton-mass spectrum and completely obscure the interesting
underlying short-distance physics at intermediate masses. 
It is evident that appropriate cuts
have to be applied in the mass-spectrum, but they cannot completely
eliminate the resonances' tails that will show up in the measured
value of the Wilson-coefficient $C_9$. One thus defines a (process-
and momentum-dependent) ``effective'' Wilson-coefficient, $C_9^{\rm
  eff}(s) = C_9 + Y(s)$, where $s$ is the dilepton invariant mass and
$Y$ describes long-distance effects associated with the $c\bar c$
loop. Sometimes $Y(s)$ is written as sum of ``perturbative'' and
``resonance'' contributions, where the former comprise just the
perturbative loop-diagrams (to leading order in $\alpha_s$; explicit
O$(\alpha_s)$ corrections are not yet known), and the latter are
expressed as sum over Breit-Wigner amplitudes. 
Actually, however, there is no such
clear-cut separation of resonance and perturbative contributions; they
are, on the contrary, dual to each other in the sense that the
perturbative result, valid well below the threshold of $J/\psi$
production, should match the dispersion integral over the resonance
region; summing up both contributions leads to double-counting. 
A clear discussion of this issue
can be found in~\cite{KS}, where $Y(s)$ was calculated from all available
information on resonances using the factorization approximation, i.e.\
neglecting gluon-exchange between the $c\bar c$ pair and the $b$ and
$s$ quark. A calculation of the dominant nonperturbative (soft-gluon)
terms from operator product expansion proves feasible at nonzero $s$,
well below the $J/\psi$ threshold, and yields,
analogous to $b\to s\gamma$, 
contributions suppressed as $1/m_c^2$.$\,$\cite{BI} A corresponding
calculation for the exclusive case is still missing.

In summary, the measurement of Wilson-coefficients in $b\to
s\ell^+\ell^-$ transitions is, even well below the $c\bar c$
resonances in the dilepton-mass spectrum, affected by long-distance
contributions that still need to be assessed in more detail.

\section{Summary \& Conclusions}

Summarizing, I have given a status report of currently much heeded
topics in B physics related to B-mixing and QCD effects in
CP-asymmetries and rare decays, whose understanding is essential for a
theoretically clean extraction of CP-violating phases and new-physics
effects from experimental results. I am confident that by the time of
the next Rencontres de Vietnam the fruitful mutual interaction of
experimental and theoretical developments will have resulted in a
much better comprehension of the mechanism underlying observable
CP-violating effects and -- maybe -- even have led to
unequivocal evidence for new physics.

\section*{Acknowledgments}
I would like to thank the organizers for inviting me to participate in
this stimulating event and DFG, the German Research Council, for
financial support. And I apologize to all people whose papers are
missing in the list of references -- 
space limitations forced me to often refer to review articles
rather than give a comprehensive list of original papers.

\end{document}